\documentclass{mn2e}

\usepackage{amsfonts}
\usepackage{amsmath}
\usepackage{graphicx}

\title[The halo-model description of marked statistics]
      {The halo-model description of marked statistics}
\author[R. K. Sheth]
{Ravi K. Sheth\thanks{E-mail: shethrk@physics.upenn.edu}\\
 Department of Physics \& Astronomy, University of Pennsylvania, 
      PA 19104, USA}

\newcommand{\bm}[1]{{\mbox{\boldmath $#1$}}}

\begin{document}

\pagerange{\pageref{firstpage}--\pageref{lastpage}}

\maketitle
\label{firstpage}

\begin{abstract}
Marked statistics allow sensitive tests of how galaxy properties 
correlate with environment, as well as of how correlations 
between galaxy properties are affected by environment.  
A halo-model description of marked correlations is developed, 
which incorporates the effects which arise from the facts that 
typical galaxy marks (e.g., luminosity, color, star formation rate, 
stellar mass) depend on the mass of the parent halo, and that massive 
haloes extend to larger radii and populate denser regions.  
Comparison with measured marked statistics in semi-analytic galaxy 
formation models shows good agreement on scales smaller than a 
Megaparsec, and excellent agreement on larger scales.  
The halo-model description shows clearly that the behaviour of some 
low-order marked statistics on these scales encodes information about 
the mean galaxy mark as a function of halo mass, but is insensitive 
to mark-gradients within haloes.  
Higher-order statistics encode information about higher order moments 
of the distribution of marks within haloes.  
This information is obtained without ever having to identify haloes 
or clusters in the galaxy distribution.  
On scales smaller than a Megaparsec, the halo-model calculation shows 
that marked statistics allow sensitive tests of whether or not central 
galaxies in haloes are a special population.  
A prescription for including more general mark-gradients in the 
halo-model description is also provided.  The formalism developed 
here is particularly well-suited to interpretation of marked 
statistics in astrophysical datasets, because it is phrased in the 
same language that is currently used to interpret more standard 
measures of galaxy clustering.  
\end{abstract}



\section{Introduction}
Almost all clustering analyses to date treat galaxies as points 
without attributes.  However, galaxies have luminosities, sizes, 
shapes, velocity dispersions, star formation rates, etc.  Recent 
work (Hamilton 1988; Norberg et al. 2002; Zehavi et al. 2005) has 
begun to study how galaxy correlations depend on luminosity and 
color---the more luminous galaxies are more strongly clustered, 
and red galaxies tend to cluster more strongly than blue.  
However, the quality of the data is now sufficiently good that one 
can imagine measuring, not just galaxy clustering as a function of 
galaxy attribute, but the spatial correlations of the attributes 
themselves.  That is to say, rather than measuring clustering as a 
function of luminosity, one can now measure the clustering of luminosity 
(or of color, star-formation rate etc.).  

B\"orner, Mo \& Zhao (1989) were among early pioneers, studying the 
correlation functions of galaxies with different weightings according 
to luminosity and mass.  Although also discussed by Peebles (1980), this 
sort of approach has been formalized under the framework of marked 
point processes (e.g. Stoyan 1984; Stoyan \& Stoyan 1994).  
Marked statistics have recently been applied to astrophysical datasets 
by Beisbart \& Kerscher (2000), Beisbart, Kerscher \& Mecke (2002), 
Gottl\"ober et al. (2002) and Faltenbacher et al. (2002). 

Marked statistics provide a useful framework for describing point processes 
in which the points have attributes or weights.  They are particularly 
well-suited to identifying and quantifying correlations between galaxy 
properties (luminosities, colors,, stellar masses, star formation rates) 
and their environments (e.g. Sheth, Connolly \& Skibba 2005), 
particularly when such correlations are weak (Sheth \& Tormen 2004).  
The halo model (reviewed in Cooray \& Sheth 2002) is the framework 
within which traditional (i.e. unmarked) measurements of galaxy 
clustering are currently interpretted (e.g., Magliochetti \& Porciani 2003, 
Mo et al. 2004; Zehavi et al. 2005; Collister \& Lahav 2005).  
This paper develops the halo-model description of marked statistics.  

Section~\ref{define} defines a number of marked statistics.  
Section~\ref{model} provides a halo model calculation of these marked 
statistics, under the assumption that marks do not correlate with 
spatial position within haloes.  
The analysis extends ideas presented in Sheth, Abbas \& Skibba (2004).  
It then compares the halo-model description with measurements in 
simulations.  
Section~\ref{markgrads} shows how the halo-model description can 
be extended to allow for correlations between marks and position 
within halo---mark gradients.  It pays special attention to the 
case in which the central object in a halo is different from all 
the others.  
The analysis shows how the halo model can be used to test simple 
physical models of why galaxy properties correlate with environment.  
On larger scales, the halo-model description of marked statistics 
shows that they can be thought of as being linearly biased versions 
of unweighted statistics---this is the subject of Section~\ref{biasing}.  
A final section discusses how the methods presented here provide 
the basis for interpretating measurements of marked statistics which 
can be made with databases currently available.  It also shows how 
marked statistics can be used to interpret measurements which indicate 
that correlations between galaxy properties (e.g., the correlation 
between stellar mass and $K$-band luminosity) also correlate with 
environment.  
An Appendix illustrates some of the key ideas using a fully 
analytic toy model.

\section{Marked statistics}\label{define}
In what follows, a mark is a weight or attribute associated with each 
point in a point process.  To make the discussion less abstract, we 
will often use astrophysical terms to illustrate our arguments.  
Thus, a point process is a galaxy catalog, and a mark can be any 
observable property associated with a galaxy, such as luminosity, 
color, velocity dispersion, size, star formation rate, etc.  
Marked statistics measure the clustering of marks.  
Since the positions at which the marks are measured may themselves 
be clustered, marked statistics are defined in a way which accounts 
for this.  

For example, let $\bar\rho$ denote the mean density of particles, 
and let $\bar w$ denote the mean mark, averaged over all particles.  
Now consider a particle with mark larger than this mean value.  
Are the particles neighbouring it also likely to have larger marks?  
One way to quantify this is to compute the ratio of the mean mark 
to $\bar w$ of pairs of particles as a function of pair separation.  
The typical number of pairs at separation $r$ is 
$\bar\rho^2 [1 + \xi(r)]$, where $\xi$ is the two point correlation 
function.  Therefore, the mean mark is 
\begin{eqnarray}
 {\cal M}_1(r) &=& {\sum  [w({\bm x}) + w({\bm y})]\,
                          {\cal I}(|{\bm x}-{\bm y}|-r) 
                  \over 2\bar w\ \sum {\cal I}(|{\bm x}-{\bm y}|-r)} 
                  \nonumber\\
               &=& {\sum  [w({\bm x}) + w({\bm y})]\,
                          {\cal I}(|{\bm x}-{\bm y}|-r) 
                  \over 2\bar w\ \bar\rho^2 [1+\xi(r)]} ,
 \label{M1}
\end{eqnarray}
where ${\cal I}(x)=0$ unless $x=0$, and the sum is over all galaxy 
pairs.  We have divided by $\bar w$, so ${\cal M}_1(r)=1$ for all $r$ 
if there are no correlations between marks.  
Analogously, the $n$th-order mark is defined by 
\begin{equation}
 {\cal M}_n(r) =  {\sum [w({\bm x}) + w({\bm y})]^n\,
                          {\cal I}(|{\bm x}-{\bm y}|-r) 
                   \over (2\bar w)^n\ \bar\rho^2 [1+\xi(r)]}.
 \label{Mn}
\end{equation}
For what follows, it is also useful to define 
\begin{equation}
 {\cal C}_n(r) =  {\sum [w({\bm x}) - w({\bm y})]^n\,
                          {\cal I}(|{\bm x}-{\bm y}|-r) 
                   \over (2\bar w)^n\ \bar\rho^2 [1+\xi(r)]}.
 \label{Cn}
\end{equation}
It is sufficiently straightforward to generalize these concepts of 
$n$-th order marks from pairs to $N$-tuples that we have not written 
the expressions explicitly.  

In what follows, we will mainly study the cases when $n=1$ and 2.  
In this regard, it is helpful to re-write ${\cal M}_2$ as 
\begin{eqnarray}
 {\cal M}_2(r) &=& {\sum [w({\bm x}) - w({\bm y})]^2\  
                          {\cal I}(|{\bm x}-{\bm y}|-r) 
                   \over (2\bar w)^2\ \bar\rho^2 [1+\xi(r)]}\nonumber\\
       && \quad + \quad 
                   {\sum  w({\bm x})w({\bm y})\,{\cal I}(|{\bm x}-{\bm y}|-r) 
                   \over \bar w^2\ \bar\rho^2 [1+\xi(r)]}.  
 \label{M2}
\end{eqnarray}
The second term involves a similar sum to that which defines $1+\xi$, 
except that now each particle of the pair contributes a weight $w$.  
Thus, the second term can be thought of as a `weighted' correlation 
function.  If we write it as $[1+W(r)]/[1+\xi(r)]$ then 
\begin{equation}
 {\cal M}_2(r) = {\cal C}_2(r) + {1+W(r)\over 1+\xi(r)}.  
 \label{M2C2}
\end{equation}
When $n=2$, it is perhaps more intuitive to study the mark variance 
and covariance, defined by  
\begin{equation}
 {\rm var}(r) = {\cal M}_2(r) - {\cal M}_1^2(r) + {\cal C}_2(r)
\end{equation}
and 
\begin{equation}
 {\rm cov}(r) = {\cal M}_2(r) - {\cal M}_1^2(r) - {\cal C}_2(r).
\end{equation}

To help build intuition, it is perhaps useful to consider how one 
might estimate these marked statistics in a data set.  Consider the 
quantity $1 + \xi(r)$ which appears in the denominator of all the 
expressions above.  A common estimator for it is to simply sum the 
number of data pairs with separation $r$ in the point distribution 
and divide it by the number of pairs of similar separation in a random 
distribution.  The suggestive notation for this estimator is $DD/RR$.  
Now consider the quantity $(1+W)/(1+\xi)$.  
Similarly suggestive notation for the estimator of $1+W(r)$ is $WW/RR$.  
However, since we are interested in the ratio of these two terms, the 
appropriate estimator is $WW/DD$.  Note that $DD$ is precisely the 
term in the sum in the denominator of equation~(\ref{M1}).  
Since $WW/DD$ is simply the average over all pairs in the sample of 
the product of the weights, it can be estimated without explicitly 
constructing a random catalog, and without explicitly worrying about 
the survey geometry.  Similarly simple estimators for the other 
marked statistics defined above can also be constructed (e.g., 
${\cal M}_1(r)$ can be estimated as $WD/DD$), making them far less 
time-consuming to estimate than the usual unweighted statistics such 
as $\xi$.  

\section{The halo model description}\label{model}
This section describes how the marked statistics defined above 
can be written in the language of the halo model.  
The analysis below complements and extends ideas in 
Sheth, Abbas \& Skibba (2004).  

In the halo model (Cooray \& Sheth 2002, and references therein), 
the nonlinear density field is assumed to be made up of dense objects 
called haloes.  At any given time, haloes of different masses all have 
the same density (they are all approximately two hundred times denser 
than the background).  All mass is in such haloes, and so all galaxies 
are also associated with haloes.  

In this description, the two-point correlation function $\xi(r)$ is 
determined by the sum of two types of galaxy pairs: pairs in the same 
halo, and pairs in separate haloes.  Since the radius of a typical halo 
at $z=0$ is less than a Mpc, the one-halo term is negligible on scales 
larger than a few Mpc.  On the small scales where the one-halo term 
dominates, the shape of $\xi(r)$ is determined by how halo density 
profiles depend on halo mass, and on how halo abundances depend on 
mass; on larger scales, $\xi(r)$ is less sensitive to the shapes of 
halo profiles, and more sensitive to the clustering of the haloes 
themselves.  

For this description, it may help to think of the galaxy distribution 
as a density field, in which case 
\begin{equation}
 {\cal M}_n(r) = {\Bigl\langle[w({\bm x})+w({\bm x}+{\bm r})]^n\, 
                  \rho({\bm x})\rho({\bm x}+{\bm r})\Bigr\rangle\over 
                  (2\bar w)^n\, 
                  \Bigl\langle\rho({\bm x})\rho({\bm x}+{\bm r})\Bigr\rangle},
 \label{M1cont}
\end{equation}
where the angle brackets denote averages over all space.  

\subsection{Unweighted statistics}\label{unwtd}
In the halo model, all mass is bound up in dark matter haloes which have 
a range of masses.  Hence, the density of galaxies is 
\begin{equation}
 \bar n_{\rm gal} = \int dm \, {dn(m)\over dm}\, g_1(m),
 \label{ngal}
\end{equation}
where $dn(m)/dm$ denotes the number density of haloes of mass $m$, 
and 
\begin{equation}
 g_n(m) \equiv \sum_N N(N-1)...(N-n+1)\,p(N|m)
 \label{gnm}
\end{equation}
is  the $n$-th factorial moment of the distribution $p(N|m)$ of 
galaxies in $m$-haloes.  If $p(N|m)$ follows a Poisson distribution, 
then $g_n(m)=g_1^n(m)$.  

The correlation function is the Fourier transform of the power spectrum 
$P(k)$:
\begin{equation}
 \xi(r) = \int {dk\over k}\, {k^3 P(k)\over 2\pi^2}\, {\sin kr\over kr} .
 \label{xir}
\end{equation}
In the halo model, $P(k)$ is written as the sum of two terms: 
one that arises from particles within the same halo and dominates 
on small scales (the 1-halo term), 
and the other from particles in different haloes which dominates 
on larger scales (the 2-halo term).  Namely, 
\begin{equation}
 P(k) = P_{1h}(k) + P_{2h}(k),
 \label{Pk1h2h}
\end{equation}
where 
\begin{eqnarray*}
 P_{1h}(k) &=& \int dm\,{dn(m)\over dm}\,
               {g_2(m)\,u(k|m)^2\over\bar n_{\rm gal}^2}, \nonumber\\
 {P_{2h}(k)\over P_{\rm Lin}(k)} &=& \left[\int dm\,{dn(m)\over dm}\,
        {g_1(m)\,u(k|m)\over \bar n_{\rm gal}}\,b(m)\right]^2 .
\end{eqnarray*}
Here $u(k|m)$ is the Fourier transform of the halo density profile 
divided by $m$, 
$b(m)$ is the bias factor which describes the strength of halo 
clustering, and $P_{\rm Lin}(k)$ is the power spectrum of the mass in 
linear theory.
When explicit calculations are made, we assume that the density profiles 
of haloes have the form described by Navarro et al. (1996), 
and that halo abundances and clustering are described by the 
parameterization of Sheth \& Tormen (1999).

\subsection{Marked statistics when marks are independent of position 
within halo}\label{wtd}
Now consider the weights.  
Let $p({\bm w}|N,{\bm r},m)\,d{\bm w}$ denote the probability that 
the $N$ galaxies at positions $({\bm r}_1,\ldots,{\bm r}_N)$ in an 
$m$-halo have weights ${\bm w} = (w_1,\ldots,w_N)$.  

As our simplest model, we will consider the case in which the 
weights do not depend on position within the halo.  
If, in addition, these weights are independent, then 
$p({\bm w}|N,m)\,d{\bm w}= \prod_{i=1}^N p(w_i|N,m)\,dw_i$.  
If the distribution of weights depends on $m$ but is independent of $N$, 
then this simplifies further to 
\begin{equation}
 p({\bm w}|N,m)\,d{\bm w} = \prod_{i=1}^N p(w_i|m)\,dw_i.
 \label{pwm}
\end{equation}  
Note that this model assumes that the weight associated with one galaxy 
is independent of the others within a halo, but that the distribution 
of weights depends on the mass of the parent halo.  
Later we will compare this model with one in which the distribution 
of weights depends on distance from the centre of the parent halo, 
but is otherwise independent of the other objects in the halo.  

The mean weight associated with galaxies in $m$-haloes is 
\begin{equation}
 \langle w|m\rangle = \prod_{i=1}^N \int dw_i\,p(w_i|m) 
                        {\sum_{i=1}^N w_i\over N} 
                    = \int dw\,p(w|m)\, w.
 \label{meanwm}
\end{equation}
The mean weight averaged over all haloes is 
\begin{equation}
 \bar w = \int dm\, {dn(m)\over dm}\,{\langle w|m\rangle\,g_1(m) 
                                      \over\bar n_{\rm gal}}\,.
 \label{meanw}
\end{equation}

If we define 
\begin{equation}
 p(w) = \int dm\, {dn(m)\over dm}\,{g_1(m)\over\bar n_{\rm gal}}\,p(w|m)
\end{equation}
then 
\begin{eqnarray}
 \langle w^n\rangle &=& \int dw\,p(w)\, w^n \nonumber\\
    &=& \int dm\,{dn(m)\over dm}\,{g_1(m)\over\bar n_{\rm gal}}
              \int dw\,p(w|m)\,w^n \nonumber\\
    &=& \int dm\,{dn(m)\over dm}\,{g_1(m)\over\bar n_{\rm gal}}\,
              \langle w^n|m\rangle.
 \label{wn}
\end{eqnarray}

The marked statistics defined in the previous section require 
averages over particle pairs.  So, for instance, 
\begin{equation}
 {\cal M}_1(r) = {1 + {\cal W}_1(r)\over 1 + \xi(r)}
\end{equation}
where ${\cal W}_1(r)$ is the Fourier transform of 
\begin{equation}
 {\cal W}_1(k) = {\cal W}_1^{1h}(k) + {\cal W}_1^{2h}(k), 
 \label{W1k1h2h}
\end{equation}
with
\begin{eqnarray*}
 {\cal W}_1^{1h}(k) &=& \int dm\,{dn(m)\over dm}
     {\langle w|m\rangle\over\bar w}\,
     {g_2(m)\,|u(k|m)|^2\over\bar n_{\rm gal}^2}, \nonumber\\
 {{\cal W}_1^{2h}(k)\over P_{\rm Lin}(k)} &=& 
         \Biggl[\int dm\,{dn(m)\over dm}\,b(m)  
                {\langle w|m\rangle\over\bar w}\,
                {g_1(m)\over\bar n_{\rm gal}}\,u(k|m)\Biggr]\nonumber\\
   & & \times\Biggl[\int dm\,{dn(m)\over dm}\,b(m)  
                {g_1(m)\over\bar n_{\rm gal}}\,u(k|m)\Biggr]
\end{eqnarray*}
and $\xi(r)$ was defined earlier.  Note the similarity between 
the integrals which define $P_{1h}$ and $P_{2h}$, and those 
for ${\cal W}_1^{1h}$ and ${\cal W}_1^{2h}$.  

Similarly, we can write 
\begin{equation}
 {\cal M}_2(r) = {[1 + \langle w^2\rangle/\bar w^2]/2 
                     + {\cal W}_2(r)\over 1 + \xi(r)}
 \label{M2W2}
\end{equation}
where ${\cal W}_2(r)$ is the Fourier transform of 
\begin{equation}
 {\cal W}_2(k) = {\cal W}_2^{1h}(k) + {\cal W}_2^{2h}(k), 
 \label{W2k1h2h}
\end{equation}
with
\begin{eqnarray*}
 {\cal W}_2^{1h}(k) &=& \int dm\,{dn(m)\over dm}
     {\langle w^2|m\rangle + \langle w|m\rangle^2\over 2\bar w^2}\nonumber\\
  &&\qquad\qquad\times\quad
     {g_2(m)\,|u(k|m)|^2\over\bar n_{\rm gal}^2}\, \nonumber\\
 {{\cal W}_2^{2h}(k)\over P_{\rm Lin}(k)} &=& 
         \Biggl[\int dm\,{dn(m)\over dm}\,b(m)  
                {\langle w|m\rangle\over \sqrt{2}\bar w}\,
                {g_1(m)\over\bar n_{\rm gal}}\,u(k|m)\Biggr]^2 \nonumber\\
        && + 
         \Biggl[\int dm\,{dn(m)\over dm}\,b(m)  
                {\langle w^2|m\rangle\over 2\bar w^2}\,
                {g_1(m)\over\bar n_{\rm gal}}\,u(k|m)\Biggr]\nonumber\\
        && \quad\times\quad \Biggl[\int dm\,{dn(m)\over dm}\,b(m)  
                      {g_1(m)\over\bar n_{\rm gal}}\,u(k|m)\Biggr],
\end{eqnarray*}
If we define the variance of the weights in $m$-haloes as 
\begin{equation}
 V^2(w|m) \equiv \langle w^2|m\rangle - \langle w|m\rangle^2,
\end{equation}
and set 
\begin{equation}
 V^2(w) \equiv \langle w^2\rangle - \langle w\rangle^2,
\end{equation}
then 
\begin{eqnarray*}
 {\cal W}_2^{1h}(k) &=& 
     \int dm\,{dn(m)\over dm}\,{\langle w|m\rangle^2\over\bar w^2}\,
     {g_2(m)\,|u(k|m)|^2\over\bar n_{\rm gal}^2}\, \nonumber\\
     && + \int dm\,{dn(m)\over dm} {V^2(w|m)\over 2\bar w^2}
     {g_2(m)\,|u(k|m)|^2\over\bar n_{\rm gal}^2}\, \nonumber\\
 {{\cal W}_2^{2h}(k)\over P_{\rm Lin}(k)} &=& 
         \Biggl[\int dm\,{dn(m)\over dm}\,b(m)  
                {\langle w|m\rangle\over \bar w}\,
                {g_1(m)\over\bar n_{\rm gal}}\,u(k|m)\Biggr]^2 \nonumber\\
        && + \Biggl[\int dm\,{dn(m)\over dm}\,b(m)  
                {\langle w^2|m\rangle\over 2\bar w^2}\,
                {g_1(m)\over\bar n_{\rm gal}}\,u(k|m)\Biggr]\nonumber\\
        && \quad\times\quad \Biggl[\int dm\,{dn(m)\over dm}\,b(m)  
                      {g_1(m)\over\bar n_{\rm gal}}\,u(k|m)\Biggr]\nonumber\\
         && - \Biggl[\int dm\,{dn(m)\over dm}\,b(m)  
                {\langle w|m\rangle\over \sqrt{2}\bar w}\,
                {g_1(m)\over\bar n_{\rm gal}}\,u(k|m)\Biggr]^2.
\end{eqnarray*}
Let $W(k)$ denote the sum of the first term of ${\cal W}_2^{1h}$ with 
the first term of ${\cal W}_2^{2h}$.  Then the Fourier transform of 
$W(k)$ is the weighted correlation function $W(r)$ 
(e.g. insert this expression in equation~\ref{M2W2} and compare with 
equation~\ref{M2C2}).  
These expressions show that ${\cal M}_1(r)$ and the weighted 
correlation function $W(r)$ encode information about the first moment 
of $p(w|m)$; information about the scatter around $\langle w|m\rangle$ 
comes from ${\cal M}_2(r)$.  

If the distribution of weights, $p(w|m)$, does not depend on $m$, 
then $\langle w|m\rangle = \bar w$, so ${\cal W}_1(k) = P(k)$, 
and ${\cal M}_1(r)=1$.  
Similarly, 
 ${\cal W}_2(k) = [(1 + \langle w^2\rangle/\bar w^2)/2]\,P(k)$, 
and so ${\cal M}_2(r) = (1 + \langle w^2\rangle/\bar w^2)/2$.  This 
is sensible:  if the distribution of weights is independent of $m$, 
then ${\cal M}_2(r)$ is simply the average value of a quantity which 
is the square of the sum of two random variates divided by $4\bar w^2$.  
Thus, if $p(w|m)$ does not depend on $m$, the marked correlations are 
constants, independent of scale, and they have the values associated 
with truly independent marks.  

However, the expressions above show that marked statistics can have 
non-trivial scale-dependence if $p(w|m)$ depends on $m$, even though 
galaxy marks do not depend on position with the parent halo, and the 
mark of one galaxy is otherwise independent of the marks associated 
with the others.  That is, galaxy marks are only correlated with the 
masses of their parent haloes; all other correlations between galaxy 
marks are a consequence of this correlation.  
In such a model, the small-scale dependence of marked correlations 
is a consequence of the fact that the size of a halo depends on its 
mass.  On larger scales, the dominant cause of nontrivial scale-dependence 
of the weighted correlation function is that the spatial distribution of 
haloes is mass-dependent.  

Notice that the two halo contribution to ${\cal M}_1(r)$ and to the 
weighted correlation function $W(r)$ depend on the combination 
$g_1(m)\,\langle w|m\rangle$; this quantity is the sum of the 
marks in a halo, averaged over all halos of mass $m$---the mean 
total mark in $m$-haloes.  
This shows that the large scale behaviour of these two statistics 
encodes information about how this quantity depends on halo mass.  
Note that this information is provided without ever actually dividing 
the galaxy distribution up into clusters.  Furthermore, if the number 
of galaxies in a halo follows a Poisson distribution, then 
$g_2(m) = g_1(m)^2$, and the one-halo contribution to $W(r)$ 
encodes information about the square of this quantity.  

\subsection{Comparison with simulations}\label{sims}
Before building a more sophisticated model, it is worth checking how 
well this simple description fares when compared with marked statistics 
measured in models of galaxy formation.  
The GIF semi-analytic galaxy formation models of 
Kauffmann et al. (1999) provide a useful testbed for the halo model 
description developed above.  Measurements of marked statistics for 
a variety of marks in these simulations have been presented in 
Sheth, Connolly \& Skibba (2005).  We study some of them here.  
In all cases, the measurements are for a sample of galaxies which 
contain more than $2\times 10^{10} h^{-1}M_\odot$ in stars at $z=0.2$ 
in a flat $\Lambda$CDM cosmology with $(\Omega_0,h,\sigma_8)=(0.3,0.7,0.9)$.  
The redshift was chosen to approximately match the median redshift of 
the 2dFGRS (Colless et al. 2001) and SDSS (York et al. 2000; 
Abazajian et al. 2003) surveys.  The mock galaxy catalog contains about 
14,665 objects in a cubical comoving volume $141h^{-1}$Mpc on a side.  

\begin{figure}
 \centering
 \vspace{-0.8cm}
 \includegraphics[width=1.1\hsize]{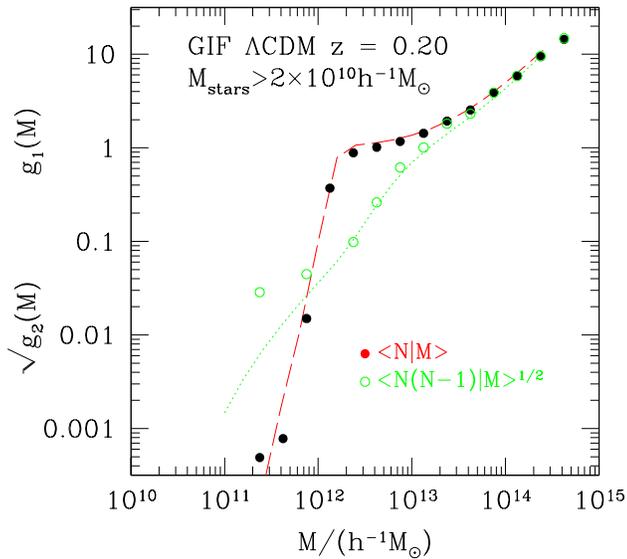}
 \caption{First and second factorial moments of the distribution of the 
          number of galaxies with stellar mass greater than 
          $2\times 10^{10}h^{-1}M_\odot$ in haloes of mass $M$ in 
          the GIF $\Lambda$CDM semi-analytic galaxy formation models 
          at $z=0.2$.  }
 \label{g1g2}
\end{figure}

The halo-model calculation requires knowledge of the first and 
second factorial moments of the galaxy counts in haloes.  
Figure~\ref{g1g2} shows how these quantities scale with halo 
mass.  The smooth curves show 
\begin{eqnarray*}
 g_1(m) &=& \left({m_{11}\over 250}\right){\rm e}^{-10/m_{11}} 
             + 1 - {\rm e}^{-(m_{11}/15)^6}
    \nonumber\\
 \sqrt{g_2(m)} &=& \left({m_{11}\over 250}\right){\rm e}^{-1/m_{11}} 
             + 0.9\, \exp\left(-{m_{11}\over 1000} - {100\over m_{11}}\right),
\end{eqnarray*}
where $m_{11}$ denotes the halo mass in units of $10^{11}h^{-1}M_\odot$.  

\begin{figure}
 \centering
 \includegraphics[width=\hsize]{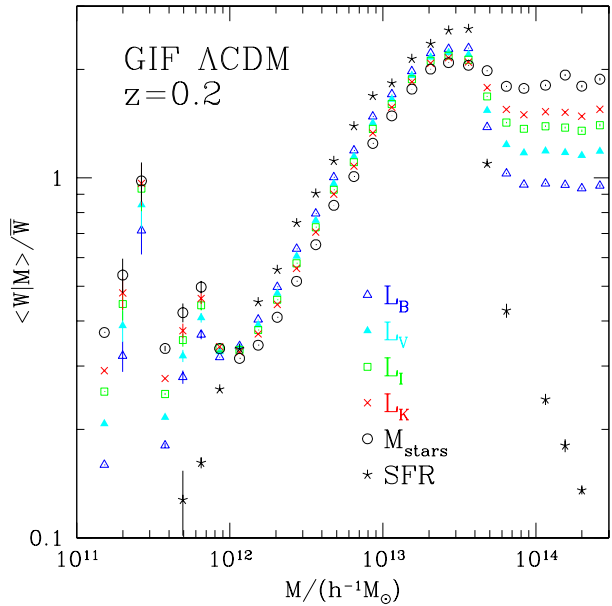}
 \caption{Mean mark as a function of parent halo mass in the GIF 
          semi-analytic galaxy formation models at $z=0.2$, for a 
          variety of marks.  While differences between the marks are 
          relatively small in the range 
          $10^{12}\le M/h^{-1}M_\odot\le 10^{13}$, more massive  
          galaxies are more luminous than the mean in the redder bands.  
          The mass dependence of any given mark gives rise to nontrivial 
          scale dependence of the associated marked statistics.}
 \label{markMhalo}
\end{figure}

In addition, calculation of marked statistics requires how the mean 
weight or mark depends on halo mass.  Figure~\ref{markMhalo} shows 
this mass-dependence for a variety of weights:  open triangles, 
filled triangles, squares, crosses, circles and stars show how the mean 
$L_B$, $L_V$, $L_I$, $L_K$, stellar mass, and star formation rate depend 
on halo mass.  Most of these marks are steeply increasing functions of
 halo mass, at least in the range 
$2\times 10^{12}\le M/h^{-1}M_\odot\le 2\times 10^{13}$.  
At larger masses, the mean luminosity-weights are approximately constant, 
but depend strongly on waveband---on average, galaxies in massive haloes 
are more luminous than average, although this over-luminosity is larger 
in the redder bands.  For a given weight, these trends with mass give 
rise to nontrivial scale dependence of the marked statistics.  
The different mass dependence of the weights makes the marked statistics 
depend on the type of mark.  

\begin{figure}
 \centering
 \vspace{-0.5cm}
 \includegraphics[width=1.05\hsize]{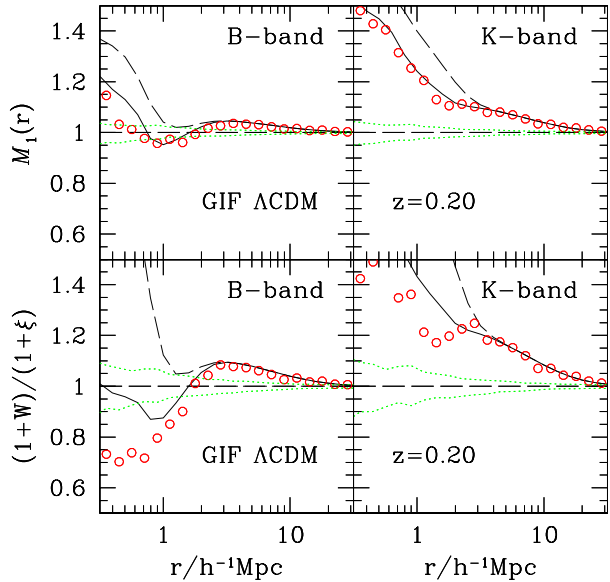}
 \caption{Marked statistics in which $B$- and $K$-band luminosities 
          were used as the mark.  
          Symbols show measurements in the GIF semi-analytic galaxy 
          formation models, and dotted curves show an estimate of 
          the uncertainty on the measurement.  Dashed curves show 
          the halo-model calculation developed earlier, in which 
          there is no distinction between the central galaxy and 
          all the others in a halo.  Solid curves show the halo-model 
          calculation described in Section~\ref{centresat}, in which central 
          galaxies are special.  }
 \label{gifLUM}
\end{figure}

Figure~\ref{gifLUM} illustrates these differences using the luminosities 
in the reddest and bluest bands as the marks.  Results for the two marked 
statistics which depend only on the mean mark within haloes are shown:  
the symbols in the top panels show ${\cal M}_1(r)$ in the simulations 
when the $B$- (left) and $K$-band (right) luminosities are used as the 
mark;  symbols in the bottom panels show the ratio of the weighted and 
unweighted pair counts $(1+W)/(1+\xi)$.  

In all panels, the dotted lines show the result of randomizing the 
marks, and then repeating the measurement of the statistic one hundred 
times.  The mean of these random realizations is shown (it is virtually 
indistinguishable from unity) bracketed by the rms scatter around it.  
This gives a rough indication of the typical uncertainty on the 
measurement (this estimated uncertainty assumes uncorrelated marks, 
so it is almost certainly an underestimate of the true error on the 
measurement).  

Note the non-trivial scale dependence of the statistics in each panel, 
and note that the scale-dependence is very different in the two bands.  
Close pairs tend to be more luminous than average in $K$, but less 
luminous than average in $B$.  

The smooth dashed curves in the different panels show the result of 
inserting $g_1(m)$, $g_2(m)$ and $\langle W|m\rangle$ from the simulations 
(c.f. Figures~\ref{g1g2} and~\ref{markMhalo}) in the halo model 
formulae given earlier.  (In practice, we approximate the two-halo 
terms using the simpler expressions given in Section~\ref{biasing}.)  
Recall that all scale dependence in these calculations is the result 
of the fact that massive haloes extend to larger radii and populate 
denser regions, and that the mean weight depends on halo mass.  
There are no additional environmental effects, and there are no 
correlations between luminosity and position with the halo.  
Comparison with the symbols shows excellent agreement on scales larger 
than $2h^{-1}$Mpc, suggesting that the analytic calculation has captured 
the essence of the physics at large separations.  
On smaller scales, however, there are differences, particularly for 
the weighted correlation functions shown in the bottom panels.  
The agreement on the larger scales which are dominated by the two-halo 
term is reassuring, because it suggests that modification to the 
one-halo term is all that is necessary to describe the statistics.  

\begin{figure}
 \centering
 \vspace{-1cm}
 \includegraphics[width=1.8\hsize]{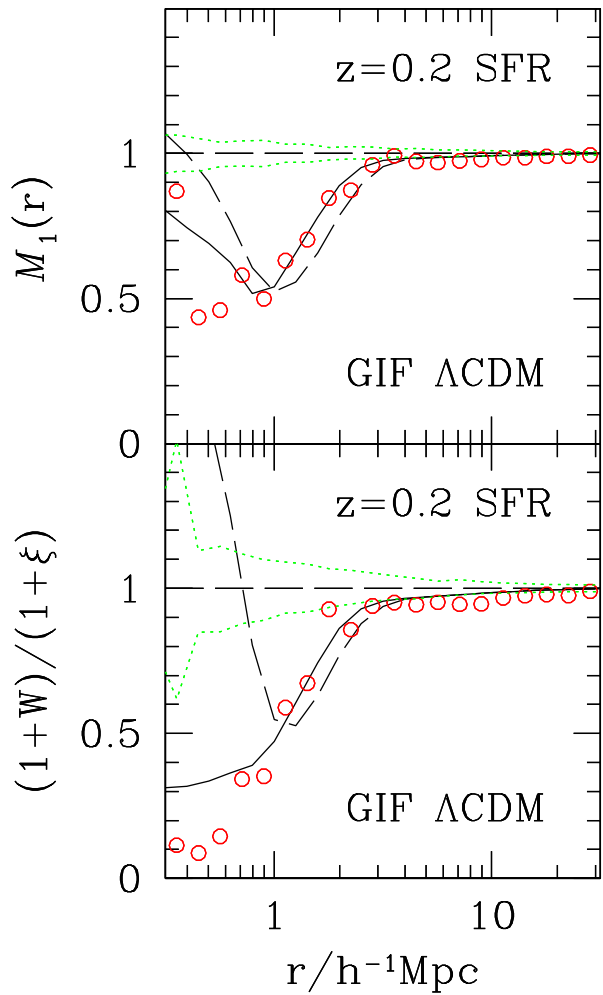}
 \caption{Marked statistics in which star formation rate was used as 
          the mark.  
          Symbols show measurements in the GIF semi-analytic galaxy 
          formation models, and dotted curves show an estimate of 
          the uncertainty on the measurement.  Dashed curves show 
          the halo-model calculation developed earlier, and 
          solid curves show the halo-model calculation developed 
          in Section~\ref{centresat}.  }
 \label{gifSFR}
\end{figure}

Figure~\ref{gifSFR} shows similar results, but now when the 
star formation rate is used as the mark.  As when the luminosity 
was the mark, the halo model calculation provides a reasonable 
description of the marked statistics on scales larger than a few Mpc, 
but it significantly over-predicts the signal on small scales.  In 
the next section, we argue that most of this discrepancy arises from 
the fact that, although the model allows for the possibility that 
marks may depend on halo mass, it does not allow marks to depend on 
position within a halo.  Thus, comparison of the model curves with the 
measurements provides some indication of the importance of such 
mark-gradients.  Evidently, such gradients only matter on small 
scales; this is sensible, since one does not expect the detailed 
distribution of marks within a halo to affect measurements on scales 
which are significantly larger than that of a typical halo.  

The next section shows how the halo model can be extended to include 
mark-gradients, thus allowing one to address the question of what 
causes these gradients.  
In particular, we show that allowing the central object in a halo 
to be different from all the others accounts for most of the 
discrepancy on small scales.  

\section{Marked statistics when marks depend on position within halo}
\label{markgrads}
This section provides two simple parametrizations of the effects of 
a correlation between galaxy mark and position within the halo.  
In the first model, this correlation is particularly simple:  
the central galaxy in a halo is supposed to be different from all 
the others, but, other than this, all the previous assumptions about 
the independence of marks apply.  This case, while simple, is a 
standard  assumption in semi-analytic and SPH-based galaxy formation 
models (e.g. Kauffmann et al. 1999; Zheng et al. 2005).  It is also 
precisely the approximation currently used to interpret measurements 
of the luminosity dependence of galaxy clustering.  
The second model allows for more sophisticated correlations between 
galaxy mark and position within the halo; it may be useful in 
studies where the mark is galaxy color, since redder galaxies in 
a halo are expected to be more centrally concentrated than the 
bluer ones.  

However, in neither model is the mark of one galaxy within a halo 
physically correlated with that of another:  the correlation is 
purely statistical.  For instance, Zheng et al. (2005) find that, 
in their semi-analytic models, there is a weak correlation between the 
number and luminosity of satellite galaxies in less massive haloes 
and the luminosity of the central galaxy:  both are smaller if the 
central galaxy is more luminous.  Such a correlation is {\em not} 
present in the models developed below.  
One signature of such a physical correlation 
would be a successful description of ${\cal M}_1(r)$ even on small 
scales, but gross discrepancies between model and measured 
$(1+W)/(1+\xi)$.  Since we see discrepancies in both the upper and 
lower panels of Figures~\ref{gifLUM} and~\ref{gifSFR}, this is less 
of an immediate concern.  In any case, accounting for this correlation 
is more complicated, and will be reported elsewhere.  

\begin{figure}
 \centering
 \includegraphics[width=\hsize]{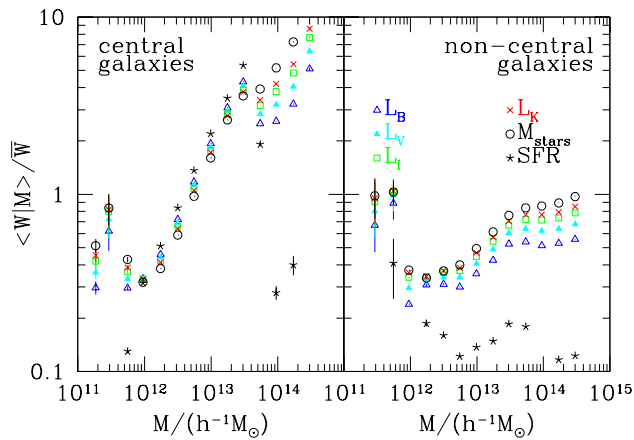}
 \vspace{-2.5cm}
 \caption{Mean mark as a function of parent halo mass in the GIF 
          semi-analytic galaxy formation models at $z=0.2$, for a 
          variety of marks.  Panel on the left shows the mean mark 
          for the central galaxy in a halo, and panel on the right 
          shows the mean mark for the other galaxies.  
          It is interesting to compare both these panels with 
          Figure~\ref{markMhalo}.  }
 \label{Mcensat}
\end{figure}

\subsection{The centre-satellite model}\label{centresat}
Figure~\ref{markMhalo} shows how the mean mark depends on halo mass.  
This mean value was computed by averaging over all the galaxies in a 
halo, whatever their location within it.  However, in the GIF models, 
the central galaxy in a halo is very different from the others.  
To illustrate, Figure~\ref{Mcensat} shows the same marks as in 
Figure~\ref{markMhalo}, but now the marks associated with the central 
galaxy (left panel) are shown separately from those associated with the 
other galaxies (right panel).  Clearly, the mass-dependence of the 
marks is very different in the two cases.  The halo model calculations 
of the previous section showed that mass dependence of any given mark 
gives rise to nontrivial scale dependence of the associated marked 
statistics.  Hence, it is possible that the failure of the halo model 
calculation on small scales (in the previous section) was due to the 
neglect of this difference between central and satellite objects.  
In this respect, the model must be extended to allow for a correlation 
between the value of the mark and its location.  

To include this effect, assume that haloes which host galaxies host 
one and only one central galaxy, and possibly many non-central 
galaxies.  We will sometimes refer to these other galaxies as satellites.  
Let $g_1^{\rm cen}(m)$ denote the fraction of $m$-haloes which host a 
central galaxy, and let $g_n^{\rm sat}(m)$ denote the $n$th factorial 
moment of the distribution of the number of satellites in a halo.  
Further, define 
\begin{equation}
 g_{1}^{\rm cs}(k|m) \equiv g_1^{\rm cen}(m) + g_1^{\rm sat}(m)\,u(k|m)\\
\end{equation}
and
\begin{equation}
 g_1^{\rm wcs}(k|m) \equiv  {g_1^{\rm cen}(m)\,\langle w_{\rm cen}|m\rangle
                    + g_1^{\rm sat}(m)\,\langle w_{\rm sat}|m\rangle\,u(k|m)
                    \over \bar w};
\end{equation}
these are the analogues of the mean number times density profile, 
and mark-weighted number times density profile.   Currently popular 
(centre plus Poisson satellite) models (e.g. Kravtsov et al. 2004) 
have 
$g_1^{\rm cen}(m)=1$ for $m$ greater than some minimum mass, 
$g_1^{\rm cen}=0$ for smaller $m$, 
$g_1^{\rm sat}(m)=0$ if $g_1(m)<1$, and 
$g_2^{\rm sat}(m)=[g_1^{\rm sat}(m)]^2$.  

Since there can only be one central galaxy, the unweighted correlation 
function $\xi(r)$ is the Fourier transform of the sum of 
\begin{eqnarray*}
 P_{1h}(k) &=& \int dm\,{dn(m)\over dm}\,
         \Biggl[{2\,g_1^{\rm cen}(m)g_1^{\rm sat}(m)\,u(k|m)\over
                 \bar n_{\rm gal}^2} \nonumber\\
   &&\qquad\qquad\qquad\qquad 
     +\ {g_2^{\rm sat}(m)\,u(k|m)^2\over\bar n_{\rm gal}^2}\Biggr], \nonumber\\
 {P_{2h}(k)\over P_{\rm Lin}(k)} &=& \left[\int dm\,{dn(m)\over dm}\,
        {g_1^{\rm cs}(k|m)\over \bar n_{\rm gal}}\,b(m)\right]^2 .
\end{eqnarray*}
The first term in $P_{1h}$ represents the contribution from 
centre-satellite--, and the second from satellite-satellite--pairs.  
(The density run of central galaxies around their host haloes is a 
delta-function.)  


Similarly, the halo model estimate of ${\cal M}_1(r)$ requires 
evaluation of 
\begin{eqnarray*}
 {\cal W}_1^{1h}(k) &=& \int dm\,{dn(m)\over dm}
     {\langle w_{\rm cen}|m\rangle + \langle w_{\rm sat}|m\rangle\over 2\bar w}
     \nonumber\\
   && \qquad\qquad \times\ 
    {2\,g_1^{\rm cen}(m)g_1^{\rm sat}(m)\,u(k|m)\over 
     \bar n_{\rm gal}^2} \nonumber\\
 &&  + \ \int dm\,{dn(m)\over dm}
   {\langle w_{\rm sat}|m\rangle\over\bar w}
   {g_2^{\rm sat}(m)\,u(k|m)^2\over \bar n_{\rm gal}^2}, \nonumber\\
 {{\cal W}_1^{2h}(k)\over P_{\rm Lin}(k)} &=& 
         \Biggl[\int dm\,{dn(m)\over dm}\,b(m)
           {g_1^{\rm wcs}(k|m)\over \bar n_{\rm gal}}\Biggr] \nonumber\\
 & & \times \Biggl[\int dm\,{dn(m)\over dm}\,b(m) \,
            {g_1^{\rm cs}(k|m)\over \bar n_{\rm gal}}\Biggr]. 
\end{eqnarray*}
And the Fourier transform of the weighted correlation function 
becomes 
\begin{eqnarray*}
 W_{1h}(k) &=& \int dm\,{dn(m)\over dm}
        {\langle w_{\rm cen}|m\rangle\over\bar w}
        {\langle w_{\rm sat}|m\rangle\over\bar w}\nonumber\\
   && \qquad \times\ {2\,g_1^{\rm cen}(m)g_1^{\rm sat}(m)\,u(k|m)\over 
                        \bar n_{\rm gal}^2}
      \nonumber\\
 && + \int dm\,{dn(m)\over dm}{\langle w_{\rm sat}|m\rangle^2\over\bar w^2} 
   {g_2^{\rm sat}(m)\,u(k|m)^2\over \bar n_{\rm gal}^2}, \nonumber\\
 {W_{2h}(k)\over P_{\rm Lin}(k)} &=& 
         \Biggl[\int dm\,{dn(m)\over dm}\,b(m)\,  
         {g_1^{\rm wcs}(k|m)\over \bar n_{\rm gal}}\,\Biggr]^2.  
\end{eqnarray*}
The solid curves in Figures~\ref{gifLUM} and~\ref{gifSFR} show these 
halo model calculations; they are in substantially better agreement 
with the measurements than the dashed curves.  (In practice, we 
approximate the two-halo terms by the simpler expressions given in 
Section~\ref{biasing}.)  

It is easy to see why this happens.  Consider, for example, the 
$K$-band luminosity.  Figure~\ref{Mcensat} shows that the central 
object is usually substantially more luminous than the satellites, 
especially at higher masses.  Moreover, the satellites are also less 
luminous than when we assign them weights in which the central object 
is not treated as special, as in Figure~\ref{markMhalo}.  
When the central object is special, then pairs with separations of 
order the diameter of a typical halo will be dominated by the 
satellite-satellite term.  Since this has smaller weights than when 
the centre was not special, the resulting values of ${\cal M}_1$ and 
$(1+W)/(1+\xi)$ are smaller.  
Similar consideration of the differences between mean satellite 
weights when the central object is and is not special explains 
the qualitative differences between the solid and dashed curves in 
Figures~\ref{gifLUM} and~\ref{gifSFR}.  

\begin{figure}
 \centering
 \vspace{-0.5cm}
 \includegraphics[width=1.05\hsize]{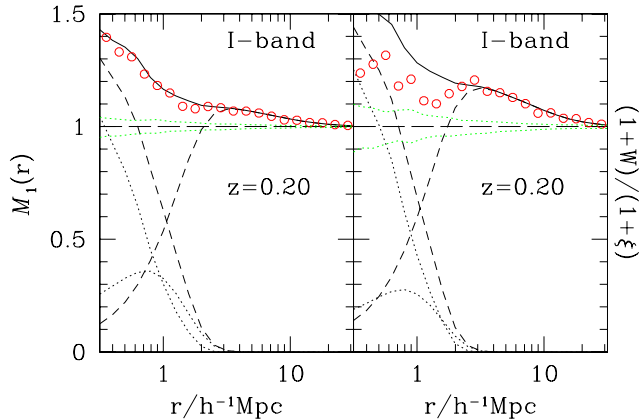}
 \vspace{-2.5cm}
 \caption{Marked statistics in which $I$-band luminosities 
          were used as the mark.  
          Symbols show measurements in the GIF semi-analytic galaxy 
          formation models, and dotted curves show an estimate of 
          the uncertainty on the measurement.  Solid curves show the 
          full halo-model calculation developed in 
          Section~\ref{centresat}, in which central galaxies are 
          special.  The two dashed curves in each panel show the one- 
          and two-halo contributions to the statistic, and the dotted 
          curves show the centre-satellite and the satellite-satellite 
          contributions to the one-halo term.  The centre-satellite 
          term dominates on the smallest scales.}
 \label{LIcs}
\end{figure}

Figure~\ref{LIcs} provides an explicit demonstration of the relative 
roles played by the various terms in the model when the mark is 
$I$-band luminosity.  The panel on the left shows results for 
${\cal M}_1$ and the panel on the right shows $(1+W)/(1+\xi)$.  
The symbols show the measured values, and the band around unity 
traced out by the dotted lines shows an estimate of the uncertainty 
on the measurements calculated by randomizing the marks (as for the 
previous figures).  The solid curve shows the full marked correlation; 
the two short dashed curves show the one- and two-halo contributions, 
and the two dotted curves show the centre-satellite (dominates on 
small scales) and satellite-satellite contributions to the one-halo 
term.  

It is worth emphasizing that, in Figures~\ref{gifLUM} and~\ref{gifSFR}, 
the mean mark in $m$-haloes is the same function of $m$ for both the 
solid and the dashed curves---the only difference is in the physical 
interpretation of this mean mark.  The solid curves represent a model 
in which the central galaxy in a halo is different from all the others, 
whereas the dashed curves show the expected marked statistics if the 
central galaxy were not special.  
Thus, our analysis shows that marked statistics are well-suited to 
discriminating between different physical models of galaxy properties.  

\begin{figure}
 \centering
 \vspace{-0.5cm}
 \includegraphics[width=1.05\hsize]{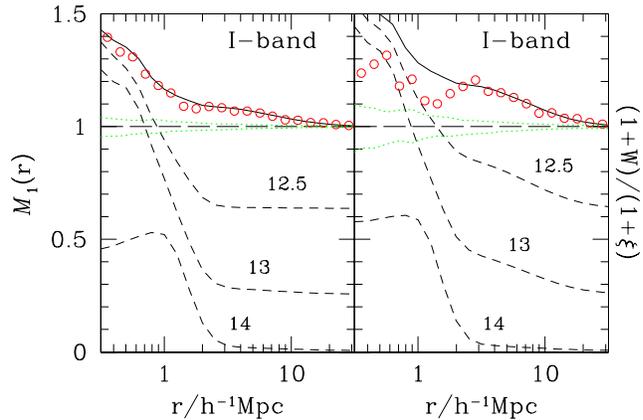}
 \vspace{-2.5cm}
 \caption{Marked statistics in which $I$-band luminosities 
          were used as the mark, shown as a function of the 
          halo mass range which contributes.  Symbols show the 
          same measured values as in the previous figure, and 
          solid curves show the same full halo-model calculation.  
          The three dashed curves in each panel show the fractional 
          contributions of the term in the numerator of the halo-model 
          expression for the marked correlation function from halos more 
          massive than $10^{14}h^{-1}M_\odot$ (bottom), 
          $10^{13}h^{-1}M_\odot$ (middle), and 
          $10^{12.5}h^{-1}M_\odot$ (top).  }
 \label{LImass}
\end{figure}

The previous plot shows how different physical models of the marks 
within halos result in different marked statistics.  
For completeness, Figure~\ref{LImass} shows how the dependence of 
the mean mark on halo mass affects the statistic.  From bottom 
to top, the different curves show the relative contribution from 
halos with masses greater than $10^{14}h^{-1}M_\odot$, 
$10^{13}h^{-1}M_\odot$, and $10^{12.5}h^{-1}M_\odot$ (meaning 
that, for the $1+\xi$ term in the denominator, the integrals were 
performed over the entire range of halo masses, but that they were 
restricted to masses greater than these values when the numerator 
was computed).  The full signal (solid curves) is very well 
approximated by the signal from halos more massive than 
$10^{12}h^{-1}M_\odot$, as one might expect from a glance at 
Figure~\ref{g1g2}.  These curves indicate that the small scale signal 
is dominated by halos with masses around $10^{13}h^{-1}M_\odot$ 
and greater, but that on larger scales, the contribution from less 
massive halos is more significant than that of more massive halos.

\subsection{A more general case}\label{grads}
This section provides a simple parametrization of the effects of 
a correlation between galaxy mark and position within the halo.  
We continue to assume that there are otherwise no correlations between 
the marks of one galaxy and another.  
Specifically, assume that there is a deterministic relation between 
distance from the halo centre and the value of the mark:  $w(r|m)$.  
(For instance, suppose some galaxy property depends on the local density 
or velocity dispersion within the parent halo.  
In reality there is almost certainly scatter in this relation; we 
think of $w(r|m)$ as an average value).  Then 
\begin{equation}
 p(w|m)\,{\rm d}w = p(r|m)\,{\rm d}r = 4\pi r^2\,{\rho(r|m)\over m}\,{\rm d}r,
\end{equation}
for some monotonic relation $w(r|m)$, and so  
\begin{equation}
 \langle w|m\rangle = \int {\rm d}w\,p(w|m)\,w 
                    = \int {\rm d}r\,p(r|m)\,w(r|m). 
\end{equation}
If we define 
\begin{equation}
 w(k|m) \equiv {\int {\rm d}r\,4\pi r^2\,w(r|m)\,\rho(r|m)\,\sin(kr)/kr \over 
                \int {\rm d}r\,4\pi r^2\,w(r|m)\,\rho(r|m)} ,
\end{equation}
then this quantity is the normalized Fourier transform of the weighted 
density profile.  

Note that although there is no scatter in the marks at fixed $r$, 
there is scatter in the marks at fixed pair separation (the two members 
of a pair of fixed separation can come from different distances from the 
halo center).  Thus, in this model, the variance in weights at fixed 
pair separation is non-trivial.  

The marked statistic ${\cal M}_1(r)$ is now given by terms like 
\begin{eqnarray*}
 {\cal W}_1^{1h}(k) &=& \int dm\,{dn(m)\over dm}
     {\langle w|m\rangle\over\bar w}\,
     {g_2(m)\,w(k|m)u(k|m)\over\bar n_{\rm gal}^2}, \nonumber\\
 {{\cal W}_1^{2h}(k)\over P_{\rm Lin}(k)} &=& 
         \Biggl[\int dm\,{dn(m)\over dm}\,b(m)  
                {\langle w|m\rangle\over\bar w}\,
                {g_1(m)\over\bar n_{\rm gal}}\,w(k|m)\Biggr]\nonumber\\
        &&\times \Biggl[\int dm\,{dn(m)\over dm}\,b(m)  
                {g_1(m)\over\bar n_{\rm gal}}\,u(k|m)\Biggr];
\end{eqnarray*}
in effect, the fact that the weight now has a profile means that one 
must replace one power of $u(k|m)$ with $w(k|m)$.  
Similarly, $W(k)$, the Fourier transform of the weighted correlation 
function, becomes $W_{1h}(k) + W_{2h}(k)$ where 
\begin{eqnarray*}
 W_{1h}(k) &=& \int dm\,{dn(m)\over dm}
     {\langle w|m\rangle^2\over\bar w^2}\,
     {g_2(m)\,|w(k|m)|^2\over\bar n_{\rm gal}^2}, \nonumber\\
 {W_{2h}(k)\over P_{\rm Lin}(k)} &=& 
         \Biggl[\int dm\,{dn(m)\over dm}\,b(m)  
                {\langle w|m\rangle\over \bar w}\,
                {g_1(m)\over\bar n_{\rm gal}}\,w(k|m)\Biggr]^2 .
\end{eqnarray*}
In this case, both powers of $u(k|m)$ have been replaced.  

Note in particular that the effect of mark-gradients is expected to 
be more dramatic for $(1+W)/(1+\xi)$ than it is for ${\cal M}_1$:  
$W_{1h}$ requires two powers of $u(k|m)\to w(k|m)$, whereas 
${\cal W}_{1h}$ only requires one.  
Thus, the analysis above indicates that incorporating weight-gradients 
in the halo-model description is relatively straightforward.  
Appendix~\ref{toy} illustrates these effects using a fully 
analytic toy model of the gradients.  We expect this model 
to be useful for studying color gradients in clusters.  

If there are true correlations between the weights of one galaxy and 
others in the same halo (such as the weak correlation reported by 
Zheng et al. 2005), then the expression for $W_{1h}(k)$ becomes more 
complicated still.  The analysis above suggests that $|w(k|m)|^2$ 
in the integrand for $W_{1h}$ should be replaced with a term which 
accounts for the correlation between the marks, as well as the shape 
of the density profile; this is the subject of work in progess.  

\section{Biasing on large scales}\label{biasing}
On scales which are larger than the diameter of a typical halo, 
the marked statistics above are dominated by the contribution 
from pairs in separate haloes.  In this limit, the scale dependence 
of marked statistics is simply related to the shape of the linear 
theory power spectrum.  To see why, consider the unweighted 
correlation function $\xi(r)$, the weighted correlation 
function $W(r)$ and the additive marked statistic ${\cal M}_1(r)$. 
On large scales, the Fourier transforms of halo profiles $u(k|m)\to 1$.  
Hence, $g_1^{\rm cs}(k|m)\to g_1^{\rm cen} + g_1^{\rm sat} \equiv g_1(m)$
and $g_1^{\rm wcs}(k|m)\to g_1(m) \langle w|m\rangle/\bar w$.  
If we define 
\begin{equation}
 b_{\rm gal} \equiv
  \int dm\,{dn(m)\over dm}\,{g_1(m)\over \bar n_{\rm gal}}\,b(m)
\end{equation}
and 
\begin{equation} 
 b_{w1} \equiv \int dm\,{dn(m)\over dm}\,{g_1(m)\over\bar n_{\rm gal}}\, 
                {\langle w|m\rangle\over\bar w}\,b(m),
\end{equation}
then
\begin{eqnarray} 
 P_{2h}(k) &\approx& b_{\rm gal}^2\,P_{\rm Lin}(k),\\
 {\cal W}_1^{2h}(k) &\approx& b_{\rm gal}\,b_{w1}\,P_{\rm Lin}(k), 
                              \qquad{\rm and}\\
 W_{2h}(k) &\approx& b_{w1}^2\,P_{\rm Lin}(k),
\end{eqnarray}
on large scales.  Thus, suitably defined combinations of $\xi$, 
$W$ and ${\cal M}_1$ provide measurements of $b_{w1}/b_{\rm gal}$.

In practice, measurement of $\xi$ requires use of a random catalog 
as well as knowledge of the survey boundary, whereas measurements of 
marked statistics do not (c.f. discussion at the end of 
Section~\ref{define}).  Hence, the most straightforward measurement 
of $b_{w1}/b_{\rm gal}$ comes from using the fact that, on large scales 
\begin{equation}
 {[1+W(r)]/[1+\xi(r)] - 1\over {\cal M}_1(r)-1} 
 \approx 1 + {b_{w1}\over b_{\rm gal}}.
\end{equation}  
The previous sections showed that, on large scales, the halo model 
calculation is in excellent agreement with the simulations.  Hence, 
our analysis shows that marked correlations allow a simple measurement 
of this ratio.  It is also straightforward to estimate the relative 
bias factors associated with two different weights:  simply measure 
$b_{w1}/b_{\rm gal}$ for each weight, and then take the ratio.

\section{Discussion}\label{discuss}
A standard assumption in semi-analytic galaxy formation models is 
that all galaxy properties are determined by the formation histories 
of their parent haloes, which, in turn, depend on halo mass.  Thus, 
correlations between galaxy properties and environment are primarily 
driven by the correlation between halo mass and environment.  It is 
these correlations which marked statistics are well-suited to 
quantifying.  The halo-model expressions for marked statistics derived 
in Section~\ref{model} have no environmental trends other than those 
which come from the dependence of halo abundances on environment.  
In essence, the halo model represents the language with which to 
describe the predictions of standard galaxy formation models.  

The extent to which simple halo-model calculations such as the 
one developed here are able to reproduce measurements of marked 
correlations in real data provides a test of the standard assumption 
that galaxy properties are more closely related to the formation 
histories of their parent haloes, rather than to additional 
environmental effects.  
Such tests are particularly interesting for two reasons.  
Recent work indicates that halo formation correlates with both mass 
{\em and} environment:  at fixed mass, haloes in dense regions formed 
earlier (Sheth \& Tormen 2004), although this effect is stronger for 
low mass haloes (Gao, Springel \& White 2005).  
One might expect to see the results of this additional environmental 
effect manifest in the galaxy distribution.  
Secondly, current halo-model based interpretations of the luminosity 
dependence of clustering (e.g. Zehavi et al. 2005) implicitly assume 
that there are no environmental effects other than those which come 
from halo biasing.  

The simplest halo model calculation (Section~\ref{wtd}) assumes that, 
within a halo, the mark associated with a galaxy is independent of 
position, and of the marks of the other galaxies in the halo.  
Despite the extreme simplicity of this model, the resulting marked 
statistics show complex scale dependence, which is entirely due to 
the fact that most galaxy attributes are strong functions of the 
masses of the haloes which host them (Figures~\ref{g1g2} 
and~\ref{markMhalo}), and halo sizes and clustering depend on halo 
mass.  

Comparison between this simple model and measurements in the GIF 
semi-analytic galaxy formation model (Figures~\ref{gifLUM} 
and~\ref{gifSFR}) indicates that the assumptions which underlie 
the halo model description are an excellent approximation on scales 
which are larger than the typical diameters of dark matter haloes.  
On these large scales, the statistics of marked pairs studied here 
can be thought of as measuring linearly biased versions of the dark 
matter power spectrum.  Prescriptions for estimating these bias factors 
are given in Section~\ref{biasing}.  

On smaller scales (those dominated by pairs in the same halo), 
the halo-model calculation which assumes that marks do not correlate 
with position within the halo is inaccurate.  This is because, in the 
GIF models, the central galaxy in a halo is different from all the 
others.  A halo-model calculation which includes this effect was 
developed in Section~\ref{markgrads}, and shown to result in 
substantially better agreement with the measurements on small scales
(Figures~\ref{gifLUM}--\ref{LImass}).  
This illustrates that marked statistics provide sharp tests of 
different physical models of galaxy formation.  
A more general model for correlations between galaxy mark and 
position within the parent halo was developed, but not tested, 
in Section~\ref{grads}, and a toy model illustrating the effects of 
mark gradients was outlined in Appendix~A.  

Galaxies are almost certainly associated with appropriately selected 
subclumps within haloes (Gao et al. 2004; Zentner et al. 2005), and so 
mark gradients are almost certainly associated with the formation 
(and tidal-stripping processes) histories of the subclumps which host 
galaxies.  Sheth \& Jain (2003) develop the formalism for incorporating 
halo substructure into the halo-model description of clustering.  
Therefore, it is likely that incorporating marked statistics into that 
formalism will prove fruitful.  Sheth, Abbas \& Skibba (2004) describe 
how to do this for the weighted correlation function---the results 
presented here show that it is straightforward to extend their analysis 
to the other marked statistics.  

\begin{figure}
 \centering
 \vspace{-1cm}
 \includegraphics[width=1.8\hsize]{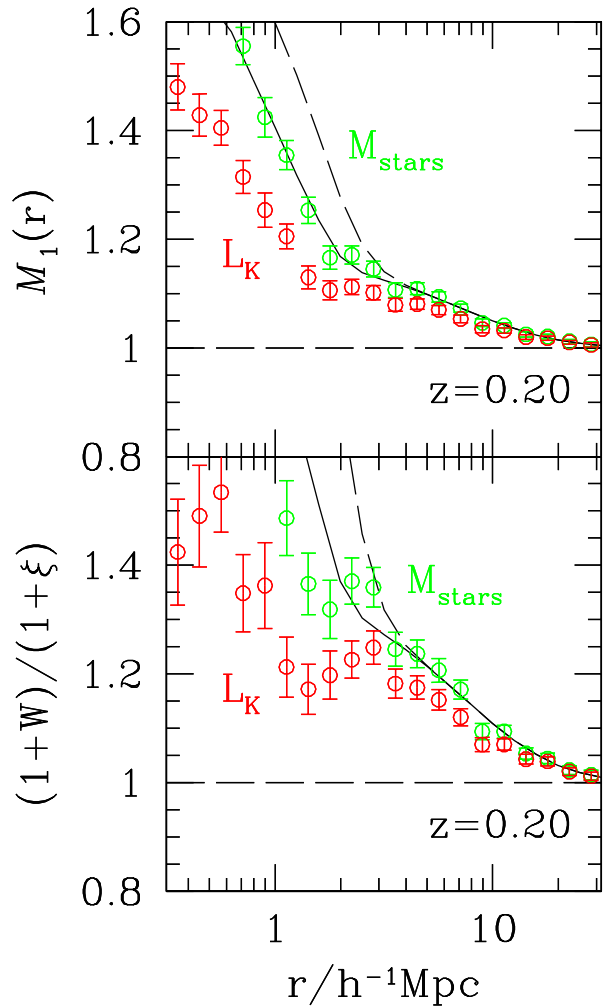}
 \caption{Marked statistics in which $K$-band luminosity and stellar mass 
          were used as marks.  
          Symbols show measurements in the GIF semi-analytic galaxy 
          formation models, and dotted curves show an estimate of 
          the uncertainty on the measurement.  Dashed and solid 
          curves show the halo-model calculation for each statistic 
          when $M_{\rm star}$ is the mark (the halo-model calculations 
          for $L_K$ are shown in Figure~\ref{gifLUM}.)  The solid 
          curves are the result of including the fact that the central 
          galaxy in a halo is different from the others.  }
 \label{gifM2L}
\end{figure}

The simplest halo model calculations presented here indicate that 
the lowest order marked statistics ${\cal M}_1(r)$ and $W(r)$ encode 
information about the mean correlation between mark and halo-mass.  
(Note that this information is provided without ever actually dividing 
the galaxy distribution up into `clusters'.)  However, this quantity 
can be estimated from other methods.  For instance, Zehavi et al. (2005) 
describe a halo-model interpretation of the luminosity dependence of 
clustering in the SDSS by studying clustering in subsamples defined 
by galaxy luminosity.  Their analysis can be used to infer how the 
mean luminosity of a galaxy correlates with the mass of its host halo, 
provided one assumes that the environment plays no additional role 
than through halo biasing.  
Therefore, insertion of their luminosity-mass correlations in the halo 
model description developed here represents a prediction for the 
shape of the luminosity-marked correlations in the SDSS.  If this 
prediction agrees with the actual measurement of marked statistics 
in the SDSS, then this will provide strong empirical justification for 
the assumption that there are no additional environmental effects.  
This test is the subject of Skibba, Sheth \& Connolly (2005).  

One may turn this statement around, and ask if the luminosity weighted 
marked statistics (${\cal M}_1$ or $W$) provide {\em any} new 
information than one gets from analysis of the luminosity dependence 
of clustering.  Clearly, marked statistics provide information about 
environmental effects which the other method does not.  However, if 
there are indeed no additional environmental effects, then our 
analysis shows that the two methods provide equivalent information 
vis a vis correlations between marks and halo masses.  
Even in this case, however, marked statistics are an attractive choice 
because they are substantially simpler to estimate (random catalogs are 
unnecessary), and they do not require division of the catalog into 
small luminosity bins to calibrate the correlation between mark and 
halo mass, thus allowing a higher signal-to-noise measurement from a 
larger catalog (rather than from smaller subsamples split by the value 
of the mark).  

Finally, Figure~\ref{gifM2L} illustrates another way in which our 
analysis aids in understanding correlations between galaxy observables 
and environment.  
The Figure compares ${\cal M}_1$ and $(1+W)/(1+\xi)$ when the mark 
is $K$-band luminosity (lower set of symbols in each panel) and when 
stellar mass (upper set of symbols) is the mark.  The trends traced 
out by both marks are the same---close pairs are more luminous and 
have larger stellar masses---although the amplitude is larger when 
stellar mass is used as the mark.  (To better show that these 
differences are significant, we have attached the error bars to each 
set of points, rather than using the same format as in the previous 
Figures.)  
The differences between the marked statistics suggest that close pairs 
have larger mass-to-light ratios.  What causes this correlation?  

Our halo-model calculation when $M_{\rm stars}$ is the mark is shown 
as the smooth curve; the analogous calculation for $L_K$ was shown 
in Figure~\ref{gifLUM}.  In both cases, the halo-model calculation 
provides an excellent description of the statistics, at least on 
scales larger than $2h^{-1}$Mpc for the weighted correlation function, 
and down to even smaller scales for ${\cal M}_1$.  
This agreement shows that, although the mass-to-light ratio is higher 
in dense regions, this environmental dependence is entirely due to the 
individual correlations between halo mass and mass-to-light ratio 
(Figure~\ref{Mcensat}), and between halo mass and environment.  
It will be interesting to see if other correlations between observables 
(e.g., the Fundamental Plane) show environmental trends for similar 
reasons.

\section*{acknowledgments}
I am grateful to the Virgo Consortium for making the data from their 
simulations available to the public at {\tt www.mpa-garching.mpg.de/Virgo}, 
the Pittsburgh Computational Astrophysics group (PiCA) for developing 
fast algorithms with which to measure correlation functions, 
Andy Connolly and Ramin Skibba for helpful conversations along the 
way, the referee Zheng Zheng for a careful reading of the manuscript, 
Sally Singh for her hospitality and Dr. Karun Singh for motivation 
during this time, and the Aspen Center for Physics for hospitality 
during the final phases of this work.  
This work is supported by NASA under grant NAG5-13270 and by the NSF 
under grant numbers AST-0307747 and AST-0520647.  


\appendix
\section{Toy model}\label{toy}
This Appendix illustrates the ideas presented in the main text using 
a toy model in which relatively transparent analytic results can be 
derived, and whose ingredients are qualitatively similar to the more 
exact calculations shown in the main text.  Let 
\begin{equation}
 \rho(r) = \rho_s\,{\exp(-r/r_s)\over (r/r_s)^2}
\end{equation}
denote the density run of the mass around a halo centre.  
The total mass associated with this profile is 
\begin{equation}
 M = \int {\rm d}r\,4\pi r^2\,\rho(r) = 4\pi\,r_s^3\rho_s
\end{equation}
The normalized Fourier transform of this profile is 
\begin{equation}
 u(k) = {1\over M}\int {\rm d}r\,4\pi r^2\,\rho(r)\,{\sin(kr)\over kr} 
      = {{\rm arctan}(kr_s)\over kr_s}.
\end{equation}
The correlation function is proportional to the Fourier transform of 
the square of this quantity.  

Now suppose that the probability a galaxy lies at distance $r$ 
from the center is 
\begin{equation}
 p(r)\,{\rm d}r = {4\pi r^2 \,{\rm d}r\,\rho(r)\over 4\pi\,r_s^3\rho_s}
                = \exp(-r/r_s)\,{{\rm d}r\over r_s};
\end{equation}
i.e., galaxies trace the mass.  
Further, suppose that objects which are more distant from the centre 
are more luminous: $L/L_* = r/r_s$ for some constant $L_*$.  
Then the distribution of luminosities is 
\begin{equation}
 p(L)\,{\rm d}L = p(r)\,{\rm d}r 
                = \exp(-L/L_*)\,{{\rm d}L\over L_*}.
\end{equation}
Thus, the mean luminosity is $L_*$.  This will be useful shortly.  
Note that both the density run $\rho(r)$ and the luminosity 
distribution $p(L)$ have rather realistic shapes, so the results 
which follow should resemble the real-world at least qualitatively.  

In this model, the run of the luminosity-weighted profile is 
\begin{equation}
 {L(r)\over L_*}\,\rho(r) = \rho_s\,{\exp(-r/r_s)\over (r/r_s)}, 
\end{equation}
so the normalized Fourier transform of this profile is 
\begin{equation}
 w(k) = {\int {\rm d}r\,4\pi r^2\,L(r)\,\rho(r)\,\sin(kr)/kr \over 
         \int {\rm d}r\,4\pi r^2\,L(r)\,\rho(r)} = {1\over 1+k^2r_s^2}.
\end{equation}
Hence, the luminosity-weighted correlation function is proportional 
to $(\pi/4)\,\exp(-r/r_s)/(2\pi^2)$.  
Since $w(k)^2/u(k)^2 \le 1$ for all $k$, this shows that 
$W(r)\le \xi(r)$ for all $r$.  Hence, in this model with luminosity 
increasing with distance from halo center, the marked correlation 
function $(1+W)/(1+\xi)$ decreases with decreasing $r$.  

Note that although there is no scatter in luminosities at fixed 
$r$, there is scatter in $L$ at fixed pair separation 
(the two members of a pair of fixed separation can come from different 
distances from the halo center).  Thus, in this model, the variance 
in weights at fixed pair separation is non-trivial.  

Now suppose that we randomize the luminosities within each halo.  
This means that the total distribution of luminosities is still 
$p(L)=\exp(-L/L_*)/L_*$, but this distribution now represents the 
probability that a galaxy in the halo has luminosity $L$ {\em whatever} 
its distance from the halo center.  In this case, the result of 
weighting each galaxy by its luminosity does not yield an $r$ 
dependent weight, so the weighted profile 
$(L/L_*)\,\rho(r) = \rho(r)$, since the mean of the weights $L$ is 
$L_*$.  Hence, in this case, $w(k)=u(k)$:  the weighted and 
unweighted correlation functions are equal.  

The calculation in the previous section assumes that there is no 
correlation between weight and position within the parent halo.  
Hence it can describe the marked statistics associated with the case 
when the luminosities have been randomized.  In this case, the 
small-scale dependence of the marked correlation function is entirely 
a consequence of the fact that the mean weight may depend on halo mass.  
If there is some correlation between weight and galaxy position within 
the halo, then this will manifest as a discrepancy between the model 
and the actual measured marked correlation.  

As a specific example of why such a discrepancy may be interesting, 
suppose that the first case (weights increase with increasing distance 
from halo center) corresponds to the luminosity distribution in a blue 
band, say $L_B$, whereas the second case corresponds to the luminosity 
in a redder band, say $L_R$.  
The color is defined as $c \equiv L_R/L_B$.  
For what follows, it is useful to define $c_* = L_{R*}/L_{B*}$.  
Since $L_B$ is a deterministic function of $r$, 
the color distribution at $r$ is due to the distribution in $L_R$:
$p(c|r)\,{\rm d}c = p(L_R=cL_B(r))\,L_B(r)\,{\rm d}c 
    = \exp[-(c/c_*) L_B(r)/L_{B*}]\,(L_B(r)/L_{B_*})\,{\rm d}c/c_*$
so the mean color at $r$ is $c_*\, L_{B*}/L_B(r) = c_*\, (r_s/r)$.  
This shows that there is a color gradient:  the halo is redder in the 
center than it is outside.  It is this color gradient which gives rise 
to the difference between the two luminosity-weighted correlation 
functions.  This illustrates that marked correlation functions encode 
information about luminosity- and hence color-gradients.  


\label{lastpage}

\end{document}